\documentclass[aps,prd,twocolumn,superscriptaddress,nofootinbib,floatfix]{revtex4-2}

\usepackage{amsmath,amssymb}
\usepackage{graphicx}
\usepackage{hyperref}
\usepackage{xcolor}
\usepackage{bm}
\usepackage{multirow}
\usepackage{booktabs}
\usepackage{placeins}
\usepackage{orcidlink}

\newenvironment{fullwidthblock}{%
  \par\onecolumngrid\vspace{2pt}%
}{%
  \par\vspace{2pt}\twocolumngrid%
}

\begin{document}

\title{Neutron Stars in Energy--Momentum Squared Gravity: Structure, Stability, and Multimessenger Constraints}

\author{Baiju Dayanandan \orcidlink{0000-0002-9911-5670}}
\email{baiju@unizwa.edu.om}
\affiliation{Natural and Medical Sciences Research Centre, University of Nizwa, Nizwa 616, Sultanate of Oman}

\author{Anirudh Pradhan
\orcidlink{0000-0002-1932-8431}}
\email{Corresponding author:pradhan.anirudh@gmail.com}
\affiliation{Centre for Cosmology, Astrophysics and Space Science,
GLA University, Mathura-281\,406, Uttar Pradesh, India}

\author{Safiqul Islam \orcidlink{0000-0003-1373-4137}}
\email{Corresponding author:sislam@kfu.edu.sa}
\affiliation{Department of Mathematics and Statistics, College of Science,
King Faisal University, P.O. Box 400, Al Ahsa 31982, Saudi Arabia}

\author{Safyan Mukhtar \orcidlink{0000-0001-5415-6762}}
\email{smahmad@kfu.edu.sa}
\affiliation{Department of Mathematics and Statistics, College of Science,
King Faisal University, P.O. Box 400, Al Ahsa 31982, Saudi Arabia}

\date{\today}

\begin{abstract}
We study nonrotating neutron stars in energy--momentum squared gravity
(EMSG). The stellar models are obtained from the modified
Tolman--Oppenheimer--Volkoff equations and four original tabulated hadronic
equations of state: WFF1, SLy4, APR4, and MPA1. We use the common coupling
set $\alpha\in\{-20,-10,0,+5,+7.5\}\,\mathrm{km}^{2}$ and construct
mass--radius and mass--central-density sequences. We compare these sequences
with the $2\,M_\odot$ mass requirement, the NICER measurements of
PSR J0030+0451 and PSR J0740+6620, and the region inferred from GW170817.
Within the sampled coupling range, the computed pre-turning-point branches
remain compatible with these four benchmarks. Negative values of $\alpha$
generally shift the high-density branches toward larger masses. Positive
values produce smaller shifts in the opposite direction. The microscopic
sound speed becomes superluminal at high density in the WFF1 and SLy4
tables, whereas APR4 and MPA1 remain causal over their tabulated ranges. For
WFF1, the causal boundary falls within a coarse table segment. Its
compatibility with the $2\,M_\odot$ requirement therefore cannot be decided
at the available density resolution. At $\alpha=-20\,\mathrm{km}^{2}$, the
effective-fluid sound-speed diagnostic becomes singular for WFF1, SLy4, and
APR4. This singularity is absent from the explicit modified TOV system. All
computed sequences remain below the standard Buchdahl reference values for
compactness and surface redshift.
\end{abstract}

\maketitle

\section{Introduction}
\label{sec:intro}

Neutron stars offer a powerful laboratory for studying cold matter at
several times nuclear saturation density. The equation of state (EOS) in
this regime is still uncertain because such densities cannot be reproduced
in terrestrial experiments. Recent observations have nevertheless narrowed
the allowed mass--radius region. NICER pulse-profile modelling has provided
radius information for PSR J0030+0451 and the massive pulsar
PSR J0740+6620
\cite{Riley:2019nicer,Miller:2019psr,Riley:2021nicer,Miller:2021radius}.
Radio-timing measurements have also confirmed neutron stars with masses
close to $2\,M_\odot$
\cite{Cromartie:2019relativistic,Fonseca:2021refined}. The binary-neutron-star
event GW170817 supplied complementary information about the EOS through its
gravitational-wave signal
\cite{Abbott:2017gw170817,Abbott:2018gw170817}. The compact secondary in
GW190814 and the low-mass compact object associated with HESS J1731$-$347
provide additional, though physically different, benchmarks for dense-matter
models \cite{Abbott:2020gw190814,Doroshenko:2022strangely}. These
observations test both the microphysics of dense matter and the theory of
gravity used to describe stellar equilibrium.

Energy--momentum squared gravity (EMSG) is a matter-coupled extension of
general relativity (GR). Its action contains a term that is quadratic in the
energy--momentum tensor. Kat\i rc\i\ and Kavuk introduced the broader
$f(R,T_{\mu\nu}T^{\mu\nu})$ framework~\cite{Katirci:2014frt}. Roshan and
Shojai then formulated the specific quadratic EMSG model
\cite{Roshan:2016energymomentum}. Board and Barrow studied its early
cosmological implications~\cite{Board:2017cosmological}. Akarsu
\emph{et al.} derived the modified stellar-equilibrium equations and used
neutron-star observations to constrain the coupling
\cite{Akarsu:2018zxl}. A recent review summarizes the theory, its matter
prescriptions, and its main applications
\cite{Cipriano:2024energymomentum}.

Quadratic and more general energy--momentum-powered corrections have been
studied in many cosmological settings. These include cosmic acceleration
\cite{Akarsu:2018energymomentum}, screening of the cosmological constant
\cite{Akarsu:2019screening}, suppression of cosmological anisotropy
\cite{Akarsu:2020screening}, and constraints from big-bang nucleosynthesis
\cite{Akarsu:2024quadratic,Jang:2025big}. Bouncing solutions have also been
examined in EMSG and related $f(\mathcal{R},T^2)$ models
\cite{Barbar:2020viability,Sharif:2024theoretical}. Low-redshift data have
been used to constrain energy--momentum-powered cosmologies
\cite{Faria:2019low}. Together, these studies show that matter-dependent
corrections can become important in high-density or high-curvature regimes.

EMSG has also been applied to charged black-hole shadows and gravitational
collapse \cite{Aliyan:2024shadow,Rudra:2024gravitational}. Stellar models
have been studied in EMSG and in related nonminimal matter--geometry
coupling theories. These investigations include general
$f(R,T_{\mu\nu}T^{\mu\nu})$ stellar configurations
\cite{Sharif:2022study}, neutron-star models with nuclear-matter input
\cite{Alam:2024impact}, hydrostatic configurations in nonminimal
geometry--matter coupling gravity~\cite{Carvalho:2020hydrostatic}, and
anisotropic or exact compact-star solutions in $T^2$-based theories
\cite{Naseer:2024anisotropic,Gul:2024impact}. These approaches use different
matter Lagrangians and stellar prescriptions. They nevertheless demonstrate
the broad interest in matter-sourced corrections to gravity at compact-star
densities.

Within EMSG, early compact-star studies considered polytropic neutron-star
models~\cite{Nari:2018compact}. Later work examined color--flavor-locked
quark stars~\cite{Singh:2021colorflavora}, anisotropic quark stars
\cite{Tangphati:2023anisotropicc}, and strongly interacting or
observationally constrained quark-matter configurations
\cite{Dayanandan:2025quarka,Banerjee:2025quarkd}. Recent studies have also
addressed universal relations for proto-neutron stars
\cite{Ghosh:2026universal} and the trace anomaly and interior curvature of
EMSG neutron stars~\cite{Swain:2026trace}. These developments motivate a
systematic study based on realistic hadronic EOS tables, a common coupling
grid, and several complementary stellar diagnostics.

In this work, we use the original tabulated WFF1, SLy4, APR4, and MPA1 EOS.
These models arise from different descriptions of dense nucleonic matter
\cite{Wiringa:1988tp,Douchin:2001sv,Akmal:1998cf,Muther:1987xaa} and cover a
useful range of stiffness. We use the original tables instead of the smooth
piecewise-polytropic representation of Ref.~\cite{Read:2009constraintsa}.
This choice preserves local features in $dP/d\rho$, which are important for
the sound-speed analysis. We integrate the modified
Tolman--Oppenheimer--Volkoff equations over the same coupling grid for all
four EOS. The resulting stellar sequences are then compared with the
observational benchmarks described above.

We examine the mass--radius relation, the first turning point of the
mass--central-density curve, the microscopic and effective-fluid sound
speeds, compactness, and surface redshift. The first turning point is used as
a practical indicator of the onset of instability. A complete radial-mode
analysis would require a separate study of the star's adiabatic response
\cite{Moustakidis:2016stability}. We compare the compactness with the
standard Buchdahl reference value~\cite{Buchdahl:1959zz}. We do not interpret
this comparison as a new universal bound on the EMSG coupling. The exterior
geometry is also important. EMSG reduces to the Schwarzschild vacuum
solution, whereas other modified-gravity models can have coupling-dependent
Schwarzschild-like exteriors. Bumblebee gravity provides one example
\cite{Casana:2017jkc}. Our aim is to identify reliable trends within the
selected coupling range and to state clearly where EOS resolution or the
effective-fluid description limits the interpretation.

The paper is organized as follows. Section~\ref{sec:theory} presents the
EMSG field equations and the modified stellar-structure equations.
Section~\ref{sec:eos_method} describes the EOS input and numerical method.
Section~\ref{sec:mr} discusses the mass--radius sequences and observational
comparisons. Section~\ref{sec:properties} presents the turning-point,
causality, compactness, and redshift results. Sections~\ref{sec:discussion}
and \ref{sec:conclusion} contain the discussion and conclusions.

\section{Stellar structure in EMSG}
\label{sec:theory}

\subsection{Field equations and effective-fluid form}

We follow the EMSG formulation of Akarsu \emph{et al.}
\cite{Akarsu:2018zxl}. The action is
\begin{equation}
S=\int\left[\frac{1}{2\kappa}(R-2\Lambda)
+\alpha T_{\mu\nu}T^{\mu\nu}+\mathcal{L}_m\right]\sqrt{-g}\,d^4x,
\label{eq:action}
\end{equation}
where $\kappa=8\pi G$, $\alpha$ is the EMSG coupling, and the metric
signature is $(-,+,+,+)$. We use units with $c=1$ in the field equations and
retain $G$ explicitly. Factors of $c$ are restored when central densities,
compactness, and surface redshift are reported. Variation of the action with
respect to the metric gives
\begin{equation}
G_{\mu\nu}+\Lambda g_{\mu\nu}
=\kappa T_{\mu\nu}
+\kappa\alpha\left(g_{\mu\nu}T_{\sigma\epsilon}T^{\sigma\epsilon}
-2\theta_{\mu\nu}\right),
\label{eq:fieldeq}
\end{equation}
where
\begin{align}
\theta_{\mu\nu}={}&-2\mathcal{L}_m
\left(T_{\mu\nu}-\tfrac12 g_{\mu\nu}T\right)-TT_{\mu\nu}
\nonumber\\
&+2T^{\gamma}{}_{\mu}T_{\nu\gamma}
-4T^{\sigma\epsilon}
\frac{\partial^2\mathcal{L}_m}
{\partial g^{\mu\nu}\partial g^{\sigma\epsilon}},
\label{eq:theta}
\end{align}
and $T=g^{\mu\nu}T_{\mu\nu}$.

For a perfect fluid,
$T_{\mu\nu}=(\rho+P)u_\mu u_\nu+Pg_{\mu\nu}$ and
$u_\mu u^\mu=-1$. Following Ref.~\cite{Akarsu:2018zxl}, we take
$\mathcal{L}_m=P$ and use the corresponding prescription for the second
variation of the matter Lagrangian. The field equations then reduce to
\begin{align}
G_{\mu\nu}+\Lambda g_{\mu\nu}={}&
\kappa\big[(\rho+P)u_\mu u_\nu+Pg_{\mu\nu}\big] \nonumber\\
&+2\kappa\alpha\rho^2
\left(1+\frac{4P}{\rho}+\frac{3P^2}{\rho^2}\right)u_\mu u_\nu
\nonumber\\
&+\kappa\alpha\rho^2
\left(1+\frac{3P^2}{\rho^2}\right)g_{\mu\nu}.
\label{eq:fieldeq-fluid}
\end{align}
These equations can be written in Einstein form by defining
\begin{align}
\rho_{\rm eff}&=\rho+\alpha(\rho^2+8\rho P+3P^2),
\label{eq:rhoeff}\\
P_{\rm eff}&=P+\alpha(\rho^2+3P^2).
\label{eq:Peff}
\end{align}
The effective energy-momentum tensor is covariantly conserved by virtue of the Bianchi identity. In contrast, the physical matter tensor is generally not conserved separately when $\alpha\neq0$.

Throughout this work, $\rho$ denotes the total energy density, not the
baryonic rest-mass density. The microscopic sound speed is therefore
$c_s^2=dP/d\rho$ in units with $c=1$. In the figures, the central energy
density is displayed as $\varepsilon_c/c^2$ in
$\mathrm{g\,cm^{-3}}$.

\subsection{Modified TOV equations}
\label{sec:tov}

We consider a static and spherically symmetric spacetime,
\begin{equation}
ds^2=-e^{2\nu(r)}dt^2+e^{2\lambda(r)}dr^2+r^2d\Omega^2,
\end{equation}
with $e^{-2\lambda}=1-2Gm(r)/r$ and $\Lambda=0$. The stellar-structure
equations are
\begin{align}
\frac{dm}{dr}={}&4\pi r^2\rho
\left[1+\alpha\rho\left(1+\frac{8P}{\rho}
+\frac{3P^2}{\rho^2}\right)\right],
\label{eq:dmdr}\\[4pt]
\frac{dP}{dr}={}&-\frac{Gm\rho}{r^2}
\left(1+\frac{P}{\rho}\right)
\left(1-\frac{2Gm}{r}\right)^{-1} \nonumber\\
&\times\left[1+\frac{4\pi r^3P}{m}
+\alpha\frac{4\pi r^3\rho^2}{m}
\left(1+\frac{3P^2}{\rho^2}\right)\right] \nonumber\\
&\times\left[1+2\alpha\rho
\left(1+\frac{3P}{\rho}\right)\right] \nonumber\\
&\times\left[1+2\alpha\rho
\left(c_s^{-2}+\frac{3P}{\rho}\right)\right]^{-1}.
\label{eq:dPdr}
\end{align}
The GR equations are recovered when $\alpha=0$. A regular stellar profile
must satisfy $dm/dr>0$, $dP/dr<0$, and $1-2Gm/r>0$. The final factor in
Eq.~\eqref{eq:dPdr} must also remain finite. The local estimate
\begin{equation}
\alpha>-[2\rho c_s^{-2}+6P]^{-1}
\label{eq:regularity-estimate}
\end{equation}
gives the relevant lower limit for negative couplings
\cite{Akarsu:2018zxl}.

The effective variables define the additional quantity
\begin{equation}
c_{s,\rm eff}^{2}\equiv\frac{dP_{\rm eff}}{d\rho_{\rm eff}}
=\frac{c_s^2+2\alpha\rho+6\alpha P c_s^2}
{1+2\alpha\rho+8\alpha P+8\alpha\rho c_s^2+6\alpha P c_s^2}.
\label{eq:cs2eff}
\end{equation}
This quantity describes the effective-fluid representation. It is not the
microscopic sound speed of the EOS. We therefore apply the causal condition
to $c_s^2=dP/d\rho$, not directly to $c_{s,\rm eff}^2$.

\subsection{Exterior spacetime and surface matching}
\label{sec:vacuum}

Every EMSG correction in Eq.~\eqref{eq:fieldeq} is constructed from the
matter tensor. These corrections vanish in vacuum, where $T_{\mu\nu}=0$.
For $\Lambda=0$, the exterior field equations therefore reduce to
$G_{\mu\nu}=0$. The exterior geometry is Schwarzschild and does not depend
on $\alpha$. For the crusted hadronic EOS used here, both $P$ and $\rho$
approach zero at the stellar surface. The interior solution is then matched
at $r=R$ to
\begin{equation}
e^{2\nu(R)}=e^{-2\lambda(R)}=1-\frac{2GM}{R}.
\end{equation}
This matching yields the standard expressions for the surface
compactness and redshift. However, a Schwarzschild exterior alone
is not sufficient to establish the Buchdahl bound, whose standard
derivation also requires suitable conditions on the stellar
interior. We therefore use $C=4/9$ and $z_s=2$ only as reference
values in Sec.~\ref{sec:compactness}.

\section{Equations of state and numerical method}
\label{sec:eos_method}

We use the WFF1~\cite{Wiringa:1988tp}, SLy4~\cite{Douchin:2001sv},
APR4~\cite{Akmal:1998cf}, and MPA1~\cite{Muther:1987xaa} equations of state.
Their original tabulated crust-plus-core forms are used instead of smooth
piecewise-polytropic fits~\cite{Read:2009constraintsa}. The four tables
contain 109, 99, 1447, and 102 pressure--density pairs, respectively. This
difference in resolution matters near a sharp causal boundary, especially
for WFF1.

For each EOS and coupling, we choose a central energy density $\rho_c$ and
the corresponding central pressure $P_c$. We impose $m(0)=0$ and integrate
Eqs.~\eqref{eq:dmdr}--\eqref{eq:dPdr} outward. The stellar surface is defined
by $P(R)=0$, and the gravitational mass is $M=m(R)$. By varying the central
density, we obtain the accessible part of the stellar sequence for each EOS
and coupling. The calculation stops if the EOS table ends before a turning
point is reached or if a regular stellar profile cannot be completed.

To compare the four EOS on the same basis, we use
\begin{equation}
\alpha\in\{-20,-10,0,+5,+7.5\}\,\mathrm{km}^2
\label{eq:alpha-grid}
\end{equation}
for every model. The most restrictive negative-coupling estimate from
Eq.~\eqref{eq:regularity-estimate} is obtained for APR4,
$\alpha_{\rm min}=-26.63\,\mathrm{km}^2$. The adopted value
$-20\,\mathrm{km}^2$ therefore remains within the regular domain of the
explicit TOV system. At $\alpha=+10\,\mathrm{km}^2$, complete profiles could
not be obtained for SLy4, APR4, and MPA1 with the present table--integration
setup. We therefore end the common positive grid at
$+7.5\,\mathrm{km}^2$. This choice is a numerical limitation of the present
calculation. It is not a physical upper bound on $\alpha$.

In the GR limit, $\alpha=0$, the maximum masses are $2.1338$, $2.0485$,
$2.1585$, and $2.4613\,M_\odot$ for WFF1, SLy4, APR4, and MPA1,
respectively. These values provide the baseline for the EMSG sequences.

\section{Mass--radius relations and observational comparison}
\label{sec:mr}

\begin{figure*}[tbp]
\centering
\includegraphics[width=\textwidth]{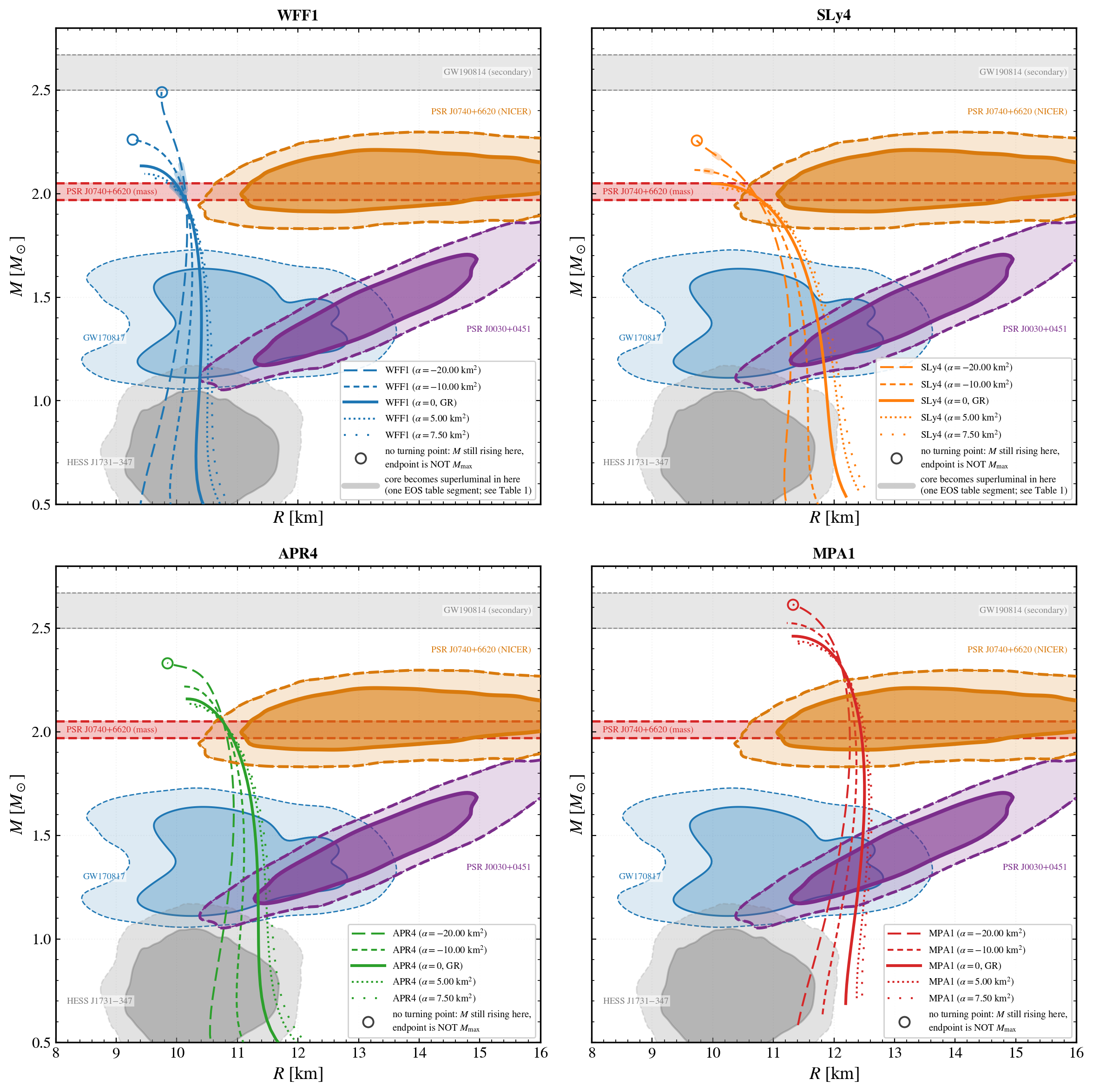}
\caption{Mass--radius sequences for WFF1, SLy4, APR4, and MPA1 over the
common coupling grid of Eq.~\eqref{eq:alpha-grid}. Solid, dashed, and dotted
curves denote $\alpha=0$, $\alpha<0$, and $\alpha>0$, respectively. Only the
pre-turning-point branches are shown. The NICER, GW170817, and pulsar-mass
regions are used for comparison. The GW190814 and HESS J1731$-$347 regions
are shown for context. Open circles mark sequences whose masses are still
increasing when the EOS table ends.}
\label{fig:mr-constraints}
\end{figure*}

Figure~\ref{fig:mr-constraints} shows the mass--radius sequences for the
common coupling grid. Negative values of $\alpha$ shift the high-density
parts of the curves toward larger masses and, in general, smaller radii.
Positive values produce smaller shifts in the opposite direction. The
coupling effect is strongest near the end of each sequence. It is weaker at
low central density because the correction is quadratic in the matter
variables.

The figure also contains six observational regions. Four are used as the
main comparison benchmarks: the $2\,M_\odot$ mass requirement, the NICER
regions for PSR J0030+0451 and PSR J0740+6620, and the region inferred from
GW170817. The secondary-mass interval of GW190814 and the HESS
J1731$-$347 region are shown only for context. Within the coupling grid of
Eq.~\eqref{eq:alpha-grid}, all computed pre-turning-point branches remain
compatible with the four adopted benchmarks. This result applies only to
the sampled models and does not provide a new observational bound on
$\alpha$.

Five sequences do not reach a mass turning point before the corresponding
EOS table ends. These are all four EOS at
$\alpha=-20\,\mathrm{km}^2$ and WFF1 at
$\alpha=-10\,\mathrm{km}^2$. The open circles in
Fig.~\ref{fig:mr-constraints} mark the final computed models. They are
computational endpoints, not values of $M_{\max}$.

\section{Turning points and physical diagnostics}
\label{sec:properties}

\subsection{Mass--central-density relation}
\label{sec:mass-epsc}

\begin{figure*}[tbp]
\centering
\includegraphics[width=\textwidth]{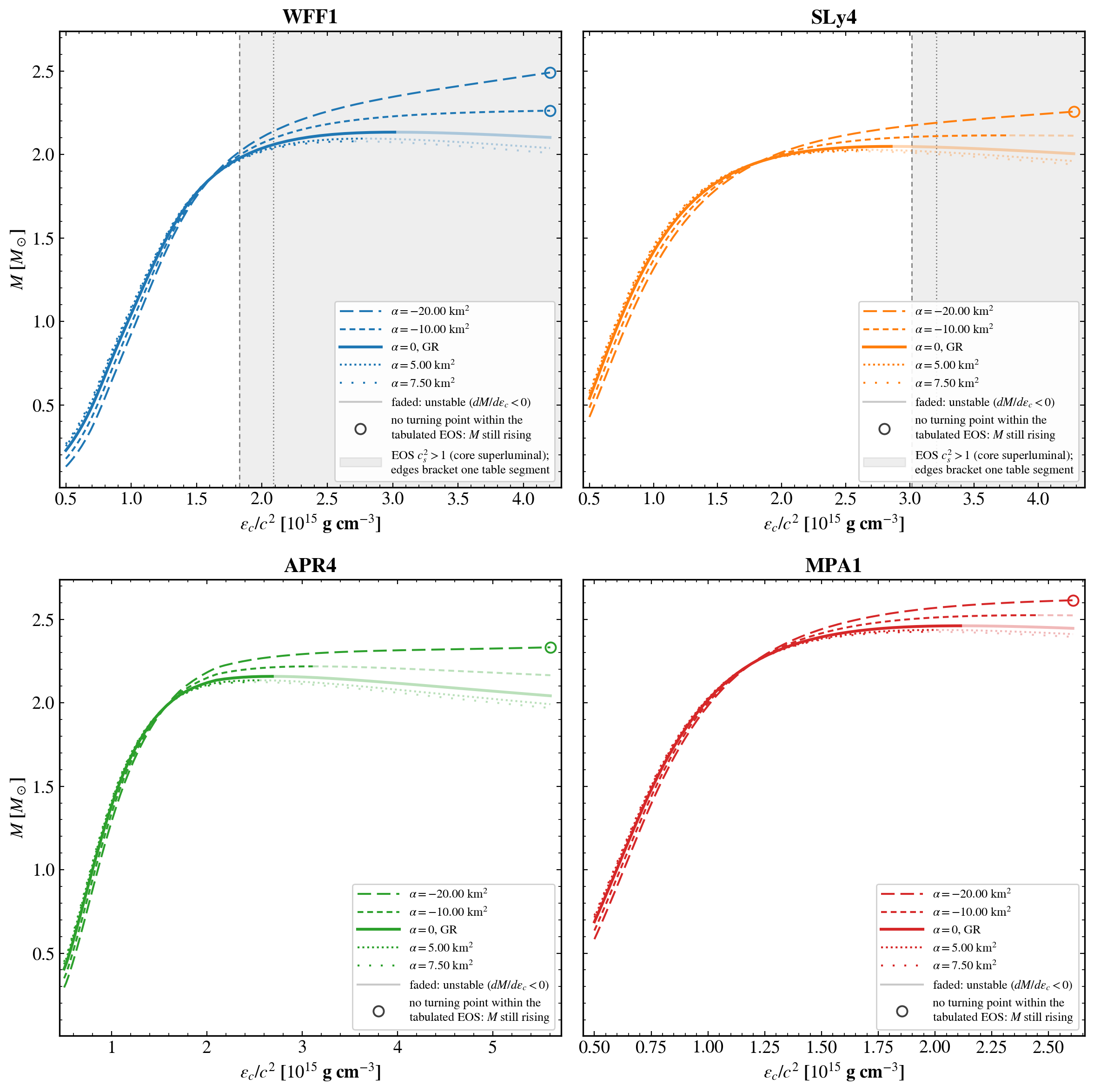}
\caption{Gravitational mass as a function of central energy density, shown
as $\varepsilon_c/c^2$. The coupling and line-style conventions are the same
as in Fig.~\ref{fig:mr-constraints}. Branches beyond the first turning point
are faded. Open circles mark sequences whose masses are still increasing at
the end of the EOS table.}
\label{fig:m-epsc}
\end{figure*}

Figure~\ref{fig:m-epsc} shows the same stellar models as functions of central
energy density. For each one-parameter sequence, we identify the first
turning point from $dM/d\varepsilon_c=0$. This point separates the
pre-turning-point branch from the later branch. It is commonly used as an
indicator of the onset of radial instability along a fixed EOS sequence.
However, it does not replace a complete radial-oscillation analysis
\cite{Moustakidis:2016stability}.

Fifteen of the twenty sequences reach a turning point within the tabulated
density range. The remaining five are the endpoint cases identified in
Sec.~\ref{sec:mr}. For these sequences, the available EOS table does not
locate the maximum mass or the change of stability.

\subsection{Sound speed and causality-limited masses}
\label{sec:soundspeed}

\begin{figure*}[tbp]
\centering
\includegraphics[width=\textwidth]{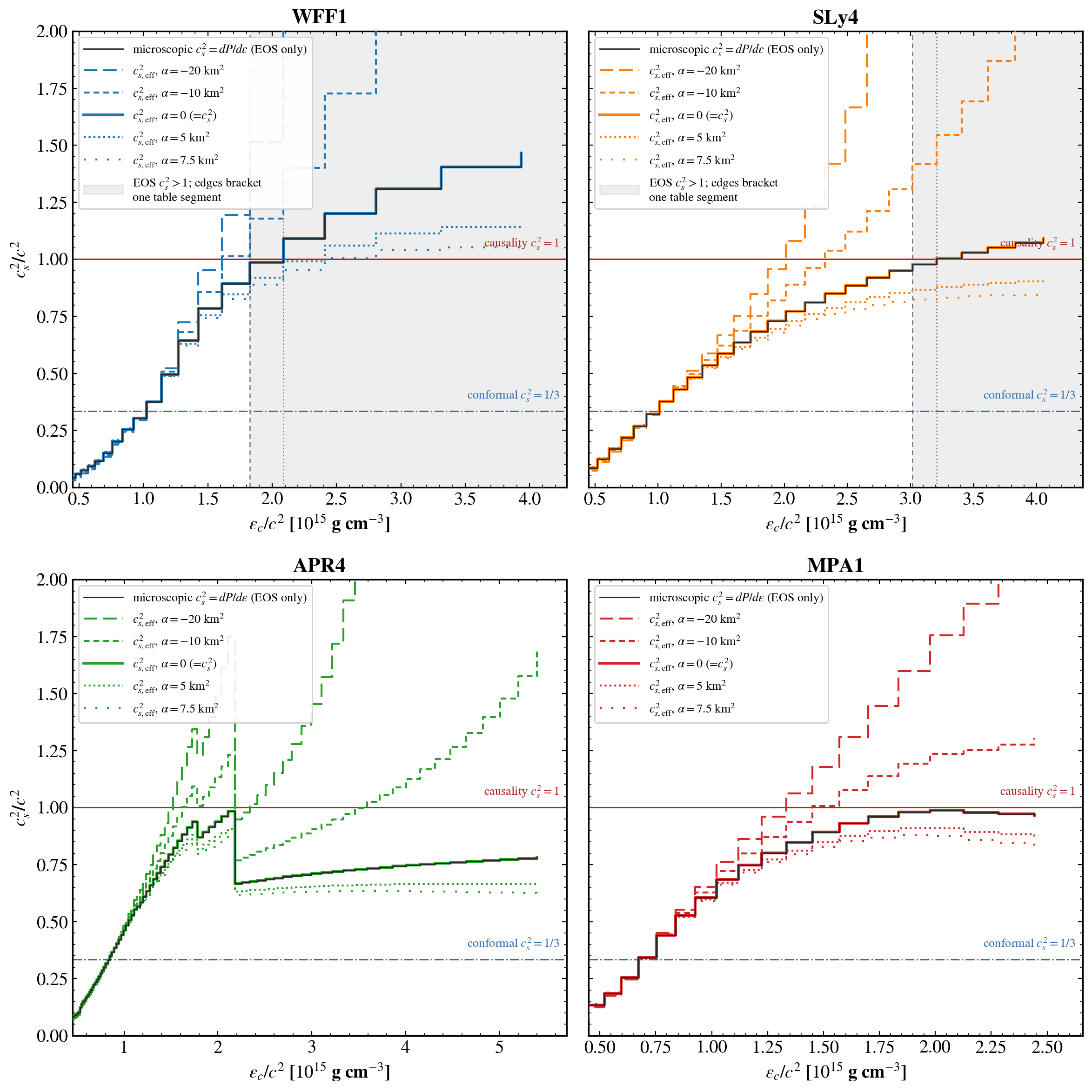}
\caption{Microscopic EOS sound speed $c_s^2=dP/d\rho$ and effective-fluid
diagnostic $c_{s,\rm eff}^2$ for the five couplings. The shaded intervals
mark the superluminal parts of WFF1 and SLy4. Values above 2 are not shown.
The vertical discontinuities in some negative-coupling curves occur when
the denominator of Eq.~\eqref{eq:cs2eff} vanishes.}
\label{fig:soundspeed}
\end{figure*}

Figure~\ref{fig:soundspeed} compares the microscopic sound speed of each EOS
with the effective-fluid diagnostic in Eq.~\eqref{eq:cs2eff}. The
microscopic sound speed does not depend on $\alpha$. In the WFF1 table, it
exceeds the causal value $c_s^2=1$ above
$\varepsilon/c^2=2.089\times10^{15}\,\mathrm{g\,cm^{-3}}$ and reaches a
maximum of 1.4839. In SLy4, it exceeds unity above
$3.206\times10^{15}\,\mathrm{g\,cm^{-3}}$ and reaches 1.1094. APR4 and MPA1
remain below unity over their tabulated ranges, with maxima of 0.9847 and
0.9888, respectively. APR4 also has a sharp softening feature near
$2.1\times10^{15}\,\mathrm{g\,cm^{-3}}$. This feature is retained because
we use the original table directly.

For WFF1 and SLy4, the turning point can occur after the first superluminal
EOS segment. Table~\ref{tab:mmax-causal} therefore also gives the mass at the
causal boundary. In WFF1, this boundary is bracketed by one coarse table
segment. The corresponding mass interval crosses $2\,M_\odot$ for every
coupling listed. The available WFF1 table therefore does not show whether
the causality-limited sequence satisfies the $2\,M_\odot$ requirement. This
ambiguity is caused by the EOS resolution, not by uncertainty in the
stellar integration.

In GR, $c_{s,\rm eff}^2$ is equal to $c_s^2$. For the positive couplings used
here, it lies below the microscopic value. For negative couplings, it lies
above it. At $\alpha=-20\,\mathrm{km}^2$, the denominator of
Eq.~\eqref{eq:cs2eff} vanishes within the density ranges sampled by WFF1,
SLy4, and APR4. The mapping from the physical variables to an ordinary
effective fluid is then not invertible. This divergence is not a singularity
of the explicit system in Eqs.~\eqref{eq:dmdr}--\eqref{eq:dPdr}. The
mass--radius profiles can still be integrated, but
$c_{s,\rm eff}^2$ should not be interpreted as a microscopic sound speed in
these intervals.

\begin{fullwidthblock}
\begingroup
\refstepcounter{table}
\label{tab:mmax-causal}
\small

\noindent\textbf{TABLE \Roman{table}.}
Turning-point or endpoint masses and the corresponding
causality-limited values for sequences extending beyond the first
superluminal EOS segment. APR4 and MPA1 are omitted because their
tabulated sound speeds remain below unity.

\vspace{0.2ex}
\centering
\setlength{\tabcolsep}{5.0pt}
\renewcommand{\arraystretch}{1.05}

\begin{tabular*}{0.96\textwidth}
{@{\extracolsep{\fill}}lccccccc}
\toprule
EOS
& $\alpha$
& $M_{\rm TP}$
& $R_{\rm TP}$
& $M_{\rm caus}$
& $R_{\rm caus}$
& $\Delta M$
& $\Delta R$
\\
&
$[\mathrm{km}^{2}]$
&
$[M_{\odot}]$
&
$[\mathrm{km}]$
&
$[M_{\odot}]$
&
$[\mathrm{km}]$
&
$[M_{\odot}]$
&
$[\mathrm{km}]$
\\
\midrule

\multicolumn{8}{l}{\emph{Turning point resolved}}\\

WFF1 & $0$ & $2.1338$ & $9.413$
& $1.978$--$2.060^{a}$ & $9.95$--$10.12$
& $0.074$--$0.156$ & $-0.54$--$-0.71$\\

WFF1 & $+5$ & $2.0953$ & $9.465$
& $1.970$--$2.043^{a}$ & $9.93$--$10.12$
& $0.052$--$0.126$ & $-0.46$--$-0.66$\\

WFF1 & $+7.5$ & $2.0789$ & $9.488$
& $1.966$--$2.035^{a}$ & $9.92$--$10.13$
& $0.044$--$0.113$ & $-0.43$--$-0.64$\\

SLy4 & $-10$ & $2.1144$ & $9.638$
& $2.1098$ & $9.884$
& $0.005$ & $-0.25$\\

\midrule

\multicolumn{8}{l}
{\emph{EOS endpoint; no turning point}}\\

WFF1 & $-20$ & $2.4910^{b}$ & $9.748$
& $2.016$--$2.140^{a}$ & $10.07$--$10.14$
& $0.351$--$0.475$ & $-0.32$--$-0.39$\\

WFF1 & $-10$ & $2.2624^{b}$ & $9.268$
& $1.996$--$2.097^{a}$ & $10.00$--$10.13$
& $0.165$--$0.267$ & $-0.73$--$-0.86$\\

SLy4 & $-20$ & $2.2567^{b}$ & $9.731$
& $2.1897$ & $10.028$
& $0.067$ & $-0.30$\\

\bottomrule
\end{tabular*}

\vspace{0.2ex}

\begin{minipage}{0.96\textwidth}
\footnotesize

\noindent
$^{a}$For WFF1, the quoted interval brackets the last causal and
first superluminal table entries and therefore reflects the EOS
resolution. The corresponding SLy4 interval is
$0.005$--$0.016\,M_{\odot}$ and is neglected.

\smallskip

\noindent
$^{b}$The EOS table ends while the mass is still increasing; the
quoted value is therefore a computational endpoint, not a maximum.

\end{minipage}
\endgroup
\end{fullwidthblock}

\subsection{Compactness and surface redshift}
\label{sec:compactness}

\begin{figure*}[tbp]
\centering
\includegraphics[width=\textwidth]{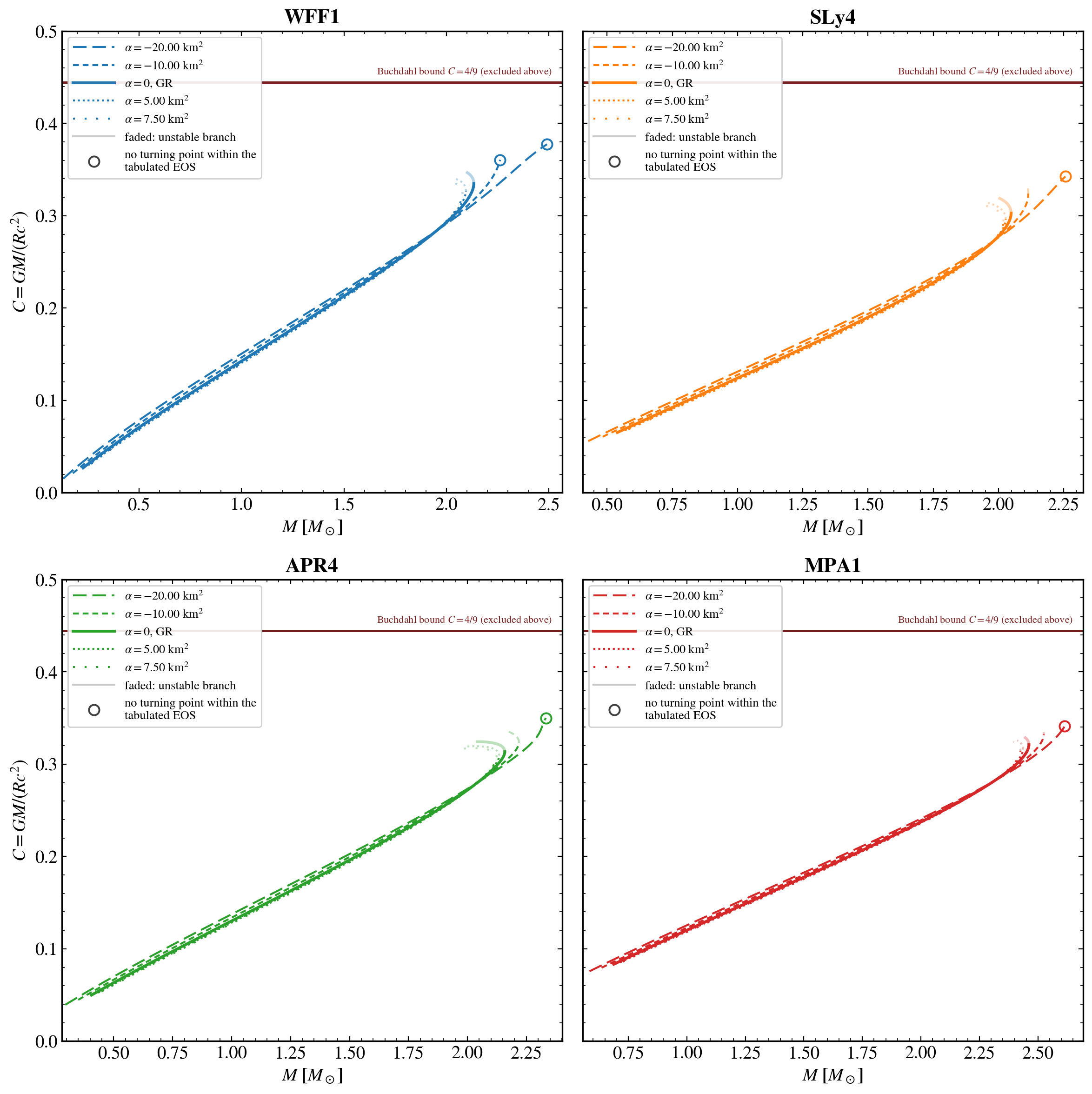}
\caption{Compactness $C=GM/(Rc^2)$ as a function of mass. The horizontal line
at $C=4/9$ is the standard Buchdahl reference. The coupling and endpoint
conventions follow Fig.~\ref{fig:mr-constraints}.}
\label{fig:compactness}
\end{figure*}

\begin{figure*}[tbp]
\centering
\includegraphics[width=\textwidth]{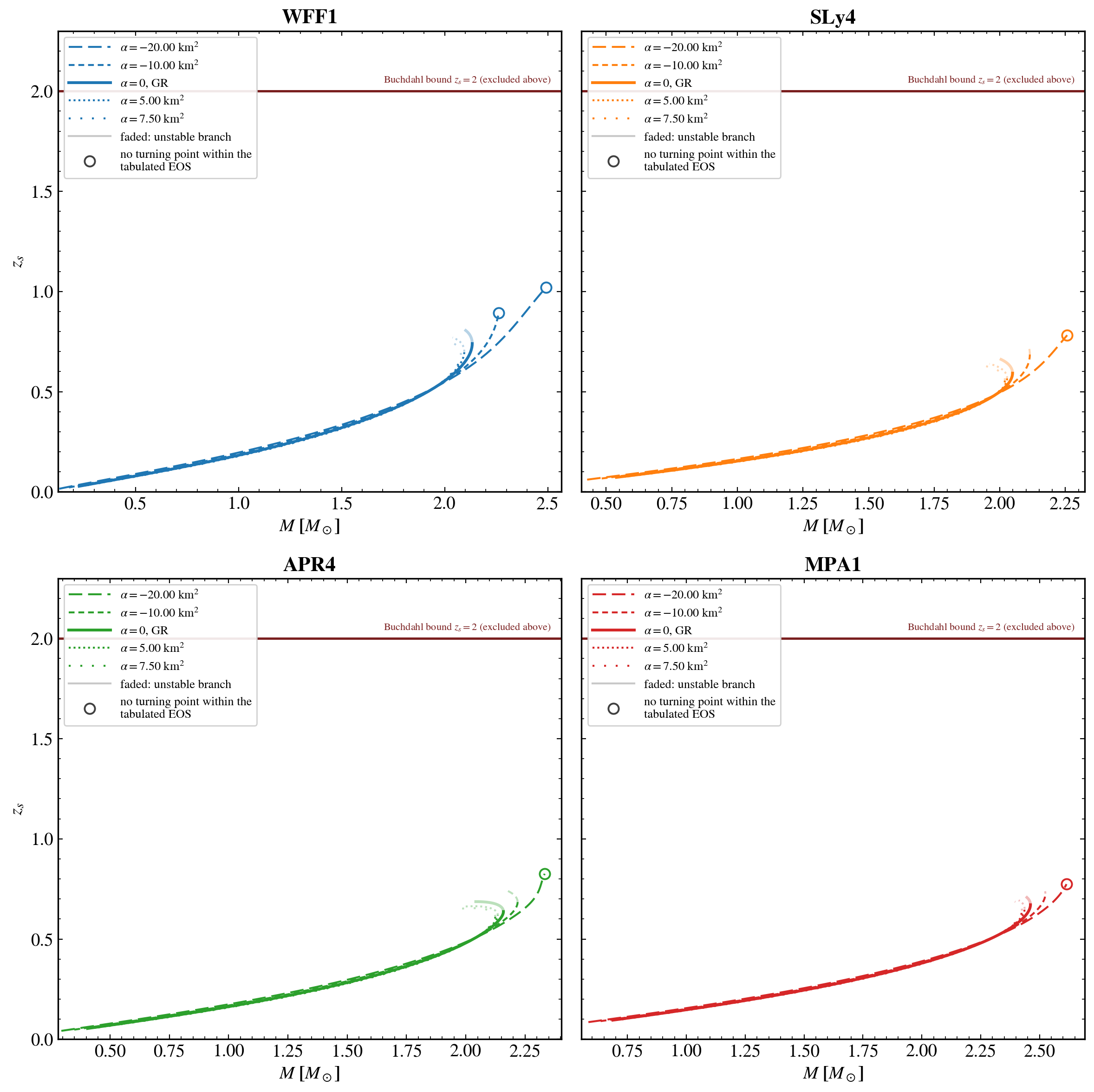}
\caption{Surface redshift $z_s=(1-2C)^{-1/2}-1$ as a function of mass. The
horizontal line at $z_s=2$ corresponds to the standard Buchdahl reference
$C=4/9$.}
\label{fig:redshift}
\end{figure*}

Figures~\ref{fig:compactness} and \ref{fig:redshift} show the compactness
$C=GM/(Rc^2)$ and the surface redshift
$z_s=(1-2C)^{-1/2}-1$. The largest compactness in the computed sample is
$C=0.3774$. It occurs for WFF1 at
$\alpha=-20\,\mathrm{km}^2$ and corresponds to a surface redshift of
$z_s=1.020$. Both values are below the standard Buchdahl references
$C=4/9$ and $z_s=2$~\cite{Buchdahl:1959zz}.

The reference lines are useful because the exterior spacetime is
Schwarzschild. However, the standard Buchdahl theorem also depends on the
assumptions made about the stellar interior. We therefore do not use the
Buchdahl line to obtain an independent constraint on $\alpha$. This caution
is especially important when the effective-fluid mapping becomes
nonmonotonic.

\clearpage

\section{Discussion}
\label{sec:discussion}

The common coupling grid shows the same qualitative trend for all four EOS.
Negative values of $\alpha$ strengthen the high-density branch and allow
larger masses within the available table range. Positive values reduce the
maximum mass by a smaller amount. The differences are weak at low density
and become clearer near the sequence endpoints. This behaviour is expected
because the EMSG correction is quadratic in the matter variables. MPA1
produces the largest masses in GR and throughout the sampled EMSG range.

The computed pre-turning-point branches remain compatible with the
observational benchmarks used in this study. Two qualifications are
important. First, the GW190814 and HESS J1731$-$347 regions are shown only
for context. Second, a turning point can occur after the microscopic EOS has
become superluminal. This issue is most important for WFF1. At the present
table resolution, its causality-limited mass cannot be placed entirely above
or below $2\,M_\odot$. A more finely sampled EOS near the causal boundary is
needed for a definite conclusion.

The calculation also has clear methodological limits. The positive coupling
grid ends at $+7.5\,\mathrm{km}^2$ because complete profiles could not be
obtained at $+10\,\mathrm{km}^2$ for three EOS. This is a numerical limit,
not an observational or theoretical bound. The first mass turning point is
used as a stability indicator, not as a replacement for a full radial-mode
calculation. In addition, the effective-fluid mapping becomes singular for
WFF1, SLy4, and APR4 at $\alpha=-20\,\mathrm{km}^2$ over part of the stellar
density range. The explicit modified TOV equations remain regular for the
reported profiles. However, these intervals should not be described as an
ordinary effective fluid.

\section{Conclusion}
\label{sec:conclusion}

We have constructed neutron-star sequences in EMSG using the original
tabulated WFF1, SLy4, APR4, and MPA1 equations of state. The same coupling
grid was used for every EOS. Negative $\alpha$ generally shifts the
high-density branch toward larger masses and compactness. Positive $\alpha$
produces smaller changes in the opposite direction. Within the sampled
grid, the computed pre-turning-point branches remain compatible with the
four observational benchmarks used for comparison.

The microscopic sound-speed analysis changes the interpretation of some
maximum masses. WFF1 and SLy4 become superluminal at high density, whereas
APR4 and MPA1 remain causal over their tabulated ranges. In WFF1, the causal
boundary is too coarsely resolved to determine whether the
causality-limited sequence satisfies the $2\,M_\odot$ requirement. This is a
limitation of the EOS table. It should not be interpreted as either support
for or exclusion of EMSG.

All computed configurations lie below the standard Buchdahl compactness and
redshift references. At $\alpha=-20\,\mathrm{km}^2$, the effective-fluid
sound-speed diagnostic becomes singular for three EOS. This does not
invalidate the explicit modified-TOV profiles, but it limits their
description as a conventional effective fluid. Stronger constraints on the
EMSG coupling will require a wider numerically stable coupling grid,
higher-resolution EOS tables with controlled causal behaviour, and a full
radial-stability analysis.

\section*{Funding}
This work was supported by the Deanship of Scientific Research, Vice
Presidency for Graduate Studies and Scientific Research, King Faisal
University, Saudi Arabia (Grant No. KFU264215).

\bibliographystyle{apsrev4-2}
\bibliography{references_refs}

\end{document}